\documentclass[a4paper,DIV=11]{scrartcl}
\usepackage[activate={true,nocompatibility},final,tracking=true,kerning=true,spacing=true,factor=1100,stretch=10,shrink=10]{microtype}
\usepackage{amsthm,amssymb,amsmath,bm,graphicx,graphbox,
  dsfont,xcolor,float,booktabs,hyperref,units,mathtools,subcaption,mycommands}
\usepackage[article]{mymath-conf}
\usepackage[Algorithm]{algorithm}
\usepackage{setspace}
\usepackage{placeins}
\usepackage{algpseudocode,tikz}
\usepackage{biblatex}
\author{Nils Margenberg
  \thanks{
    Helmut Schmidt University, Faculty of Mechanical and Civil Engineering, %
    Holstenhofweg 85, 22043 Hamburg, Germany,
    \texttt{\{margenbn,bause\}@hsu-hh.de}
  }
  \and
  Franz X. Kärtner
  \thanks{
    Center for Free Electron Laser Science (CFEL), %
    Deutsches Elektronen-Synchrotron (DESY) \& Department of Physics, University
    of Hamburg, Notkestraße 85, 22607 Hamburg, Germany,
    \texttt{franz.kaertner@desy.de}}
  \thanks{
    The Hamburg Center for Ultrafast Imaging,
    Luruper Chaussee 149, 22761 Hamburg, Germany
  }
  \and
  Markus Bause\footnotemark[1]
}
\date{\today}
\title{Accurate simulation of THz generation with Finite-Element Time Domain methods}

\begin{document}
\maketitle



\begin{abstract}
  We investigate the accurate full broadband simulation of complex nonlinear
  optical processes. A mathematical model and numerical simulation techniques in
  the time domain are developed to simulate complex nonlinear optical processes
  without the usual used slowly varying envelope approximation. We illustrate
  the accuracy by numerical simulations. Furthermore, they are used to elucidate
  THz generation in periodically poled Lithium Niobate (PPLN) including optical
  harmonic generation.
\end{abstract}


\section{Introduction}\label{sec:introduction}
Nonlinear optical phenomena are the basis of a wide range of applications such
as novel optical sources and measurement or diagnostic techniques. With the
increasing availability of high-intensity lasers, this field of research
progresses fast, continuously uncovering new applications. With growing
complexity, however, the simulation of nonlinear optical phenomena becomes more
important to achieve optimal performance and save cost and time required for
empirical studies.

In recent years wave phenomena have received increasing attention from the
applied mathematics community. Nevertheless, scientists or practitioners
oftentimes still rely on ad hoc approximations to deal with the central issues
that they face. Oftentimes these approximations allow for increased speed but
reduce the accuracy of simulations and predictions. We want to demonstrate the
accuracy and efficiency of the concepts developed in the field of applied
mathematics and apply it to problems of practical interest. In the long term, we
hope that this will reduce the need for simplifying assumptions, for example the
slowly varying envelope approximation and increase physical realism.

In this work we present efficient and accurate methods for modelling complex
phenomena in nonlinear optics. In particular we study frequency mixing processes
in the context of THz generation in periodically poled nonlinear crystals by
means of Quasi-Phase-Matching. The conceptual basis is the direct simulation of
the wave equation, which allows general combinations of dispersion models,
nonlinearities and reduced simplifying assumptions. Together with a formulation
of a perfectly matched layer (PML) for the domain truncation this allows
simulations of phenomena in nonlinear optics with high accuracy. The
discretization is done with a finite element time domain (FETD) method with
which we gain flexibility w.\,r.\,t.\ complex domain geometries, order of
approximation and better handling of discontinuities compared to e.\,g.\ finite
differences. We develop our FETD method within the framework of space time
finite element methods (STFEM). The computational effort can become enormous,
which we alleviate by using parallelization and state of the art computational
frameworks. We present numerical results for the problem of THz generation in
PPLN which demonstrates the accuracy and efficiency of the presented methods. We
conclude our work by giving an outlook for future work.

\section{Related Works}\label{sec:related-works}
Among the most popular methods to simulate electromagnetic phenomena are finite
element methods. There is a wide range of literature and software on them, which
we briefly review here. Finite element methods are often used for physical
simulations of any kind. In the field of numerical methods for electromagnetics
and optics the FETD method~\cite{jinFiniteElementMethod2014} allows fully
broadband simulations. The FETD method is well established,
cf.~\cite{jinFiniteElementMethod2014}, references therein
and~\cite{agrawalRecentTrendsComputational2017} for more recent results.
However, its use in the field of nonlinear optics results remain sparse:
  In~\cite{abrahamPerfectlyMatchedLayer2019,abrahamConvolutionFreeFiniteElementTimeDomain2019,abrahamFiniteelementTimedomainMethods2020}
    Abraham et al.\ developed methods for applications similar to the ones
    studied within this paper.
    In~\cite{abrahamParallelFiniteElementTimeDomain2020} they extended these
    methods to a mixed finite element formulation.
  In~\cite{grynkoSimulationSecondHarmonic2017} the authors use discontinuous
  Galerkin methods, which can be advantageous for hyperbolic problems due to low
  dispersion and dissipation
  errors~\cite{hesthavenNodalDiscontinuousGalerkin2008}.

In particular, no literature is available on the subject of this paper,
finite element time domain methods for the simulation of nonlinear optical
generation of THz radiation.
In~\cite{raviLimitationsTHzGeneration2014,raviTheoryTerahertzGeneration2015,wangHighEfficiencyTerahertz2018}
the authors use finite difference methods for the spatial discretization and explicit Runge-Kutta
methods for the time discretization to simulate simplified models based on the
slowly varying envelope approximation. To increase physical realism, we develop
a new approach using finite element time domain methods.
 
In the
experimental context the
generation of THz radiation is an active field of research with a wide range of
applications including imaging~\cite{guilletReviewTerahertzTomography2014},
linear and nonlinear THz spectroscopy~\cite{heblingHighpowerTHzGeneration2008}
and powering novel accelerators for compact electron sources. In the past,
research in nonlinear optical generation of THz radiation has focused on the
development of broadband, single-cycle
sources~\cite{zhangSegmentedTerahertzElectron2018,Guiramand:22}. These
techniques enabled conversion efficiencies approaching
2\%~\cite{koulouklidisObservationExtremelyEfficient2020}. Emerging applications,
such as THz-driven electron
acceleration~\cite{nanniTerahertzdrivenLinearElectron2015}, require high-field
pulses with high spectral purity. This demand has driven multiple advancements,
which improved efficiencies from the $\num{1.e-5}$ to $\num{1.e-3}$ range and
THz pulse energies from the nJ to the mJ
range~\cite{carbajoEfficientNarrowbandTerahertz2015,ahrNarrowbandTerahertzGeneration2017,jollySpectralPhaseControl2019,olgunHighlyEfficientGeneration2022}.

Electromagnetic phenomena can be complex, so simplifying assumptions such as
considering only a single frequency component (time-harmonic waves) or linear
and instantaneously responding materials are popular. Here we consider the
dispersive and nonlinear electromagnetic wave equations in the time domain. This
gives us the ability to resolve all frequencies while taking frequency dependent
responses into account.

In numerical simuations, wave propagation and other physical processes have to
be truncated to bounded regions. Artificial boundaries that do not perturb the
wave propagation and lead to unphysical reflections, caused by boundary
conditions imposed there, are desired. To this end, we discuss an implementation
of the PML, which has been introduced
in~\cite{berengerPerfectlyMatchedLayer1994}. The PML method is cost-effective,
easy to implement and not limited with respect to the computational domain. The
original PML formulation relies on a splitting of the physical field into
different components. The splitting introduces two different sets of equations
in the physical and PML domain. This requires the treatment of the interface
between the two domains. An alternative approach is to consider the PML in the
frequency domain as a coordinate
stretching~\cite{chewComplexCoordinateStretching1997}. From these formulations,
a time dependent PML can be derived. The arising convolutions can be computed by
means of auxiliary differential equations.

The PML method has been applied to lots of problems in electromagnetics and
optics~\cite{berengerPerfectlyMatchedLayer1994,chew3DPerfectlyMatched1994,sacksPerfectlyMatchedAnisotropic1995,rappaportPerfectlyMatchedAbsorbing1995,kuzuogluFrequencyDependenceConstitutive1996,wuComparisonAnisotropicPML1997,yangSpectralSimulationsElectromagnetic1997,abarbanelMathematicalAnalysisPML1997,teixeiraAdvancesTheoryPerfectly2001,collinoOptimizingPerfectlyMatched1998,collinoPerfectlyMatchedLayer1998,liuNumericalSimulationBistatic2004,abarbanelLongTimeBehavior2002,jiaoEffectiveAlgorithmImplementing2002,ramadanAuxiliaryDifferentialEquation2003,rylanderPerfectlyMatchedLayer2004,gondarenkoPerfectlyMatchedLayers2004,sjogreenPerfectlyMatchedLayers2005,louSecondorderPerfectlyMatched2007,dondericiConformalPerfectlyMatched2008,oskooiFailurePerfectlyMatched2008,fengSecondorderPMLOptimal2015,duruEfficientStablePerfectly2014,becachePerfectlyMatchedLayers2015,becacheStablePerfectlyMatched2017,becacheStablePerfectlyMatched2017,becacheAnalysisPerfectlyMatched2018,dohnalPerfectlyMatchedLayers2007}.
Here we will adapt the complex frequency shifted PML (CFS-PML). The advantage of
CFS-PML is the ability to account for a wider variety of frequency
ranges~\cite{rodenConvolutionPMLCPML2000,berengerNumericalReflectionFDTDPMLs2002,berengerApplicationCFSPML2002,correiaSimpleEfficientImplementation2005,appeloPerfectlyMatchedLayers2006,gedneyAuxiliaryDifferentialEquation2010,duruAccuracyStabilityPerfectly2012}.
This is useful for low frequency waves, where the PML is known to
fail~\cite{becachePerfectlyMatchedLayers2015}. For an in-depth review of the PML
technique we refer to~\cite{pledReviewRecentDevelopments2021} and references
therein. An additional issue in nonlinear optics is the treatment of the
nonlinear term in the PML region and at the boundary to the physical region.
In contrast to this work where the PML is left linear, Abraham et
al.~\cite{abrahamPerfectlyMatchedLayer2019} kept the PML nonlinear.

We develop the complex frequency PML in the context of space time finite element
methods (STFEM)~\cite{kocherVariationalSpaceTime2014}. This allows to use a
variety of time stepping methods, for example advantageous higher order methods
such as~\cite{anselmannNumericalStudyGalerkin2020}. Variational time
discretizations can be seen as a natural extension of variational
discretizations in space. This facilitates the use of concepts such as duality
and goal oriented
adaptivity~\cite{bauseFlexibleGoalorientedAdaptivity2021,bangerthAdaptiveGalerkinFinite2010}.
The concepts of variational space time discretization are also applied to
stability and error analysis~\cite{matthiesHigherOrderVariational2011}.
Furthermore, the use of space time FEM allow us to solve the wave equation
together with the arising auxiliary differential equations (ADE) in a single
holistic framework. This framework can then be extended by the methods mentioned
above in a generic manner.

\section{Mathematical and Physical Problem}\label{sec:mathematical-and-physical-problem}
In this section we address the physical models for the propagation of
electromagnetic waves in nonlinear dispersive media. Dispersive and nonlinear
effects are the result of the wave interacting with the atoms or molecules in a
medium. The polarization captures these interactions on a macroscopic level. The
polarization $P$ is generally developed as a power series w.\,r.\,t.\ the
electric field $E$ with the electric susceptibilities \(\chi^{(n)}\). For the
materials we consider the linear and quadratic term
\begin{equation}\label{polarization-ex}
  P=\varepsilon_{0}\paran{\chi^{(1)}E+\chi^{(2)}E^2}
\end{equation}
suffices. With the assumptions made above, this expansion of $P_{i}$ simplifies
to~\eqref{polarization-ex}. In general the coefficients \(\chi^{(n)}\) are
frequency dependent tensors of \((n+1)\)th rank. Linear dispersion can then be
described by a frequency dependent \(\chi^{(1)}\), as we will do later. While
this can also be true for the nonlinear terms, we consider instantaneous
nonlinearities in this work. With one exception, we consider homogeneous
materials. We assume that \(\chi^{(1)}\) and \(\chi^{(2)}\) can be simplified to
scalar functions. We will discuss concrete functions in later sections. Note
that, in the multidimensional case, the polarization $P$ reads as
$P_{i}=\varepsilon_{0}\paran{\chi^{(1)}_{ij}E^{j}+\chi^{(2)}_{ijk}E^{j}E^{k}}$
using Einstein summation convention. The indices range from $1$ to $d$, the
number of space dimensions.

\subsection{Physical Model}\label{sec:physical-model}
We state the governing equations in the time domain where we transform the
equations and quantites by applying a time transformation \(\tilde{t}=c_{0}t\).
We use this everywhere and omit the ``\(\,\tilde{}\,\)'' nomenclature from now
on. Let $\mathcal{D}\subset \mathbb{R}^{d}$ and $I=(0,\,T]$ be an open bounded time
interval. We state the electromagnetic wave equation that includes a second
order instantaneous nonlinearity in dispersive media:
\begin{subequations}\label{eqwave2}
  \begin{align}
    \label{eqwave2a}
    -\Delta E
    +  \partial_{tt}({{n(t)}^{2} * E})
    + \chi^{(2)}\partial_{tt}E^2&=f\quad\text{in }\mathcal{D}\times I\,,\\
    \label{eqwave2b}
    E(0) = E_{0},\quad\partial_{t}E(0)=A_{0}\quad\text{in }\mathcal{D},\quad E&=g^{E}
    \text{ on } \partial\mathcal{D}\times I\,.
  \end{align}
\end{subequations}
A Lorentz dispersion model accounts for the frequency dependent refractive
index. It models electrons as damped harmonical oscillators bound to the
nucleus. The reaction of an electron to external electromagnetic fields is then
given by the differential equation
\begin{equation}
  \label{lorentz-orig}
  m\ddot{p} +m\Gamma_{0}\dot{p} + m\nu_{t}^{2}p = -eE\,.
\end{equation}
Here \(m\) is the electron mass, \(\Gamma_{0}\) is the damping coefficient and
\(\nu_{t}\) is the phonon frequency that raises the refractive index for the low
frequencies. We use \(n_{\Omega}\) and \(n_{\omega}\) as low and high frequency
limits of the refractive index. The electric permittivity and refractive index
can then be expressed through the solution to~\eqref{lorentz-orig} above:
\begin{equation}\label{eqlorentzfreq}
  {n(\nu)}^{2}= n_{\omega}^{2}
  + \frac{(n_{\Omega}^{2}-n_{\omega}^{2})
    \nu_{t}^{2}}{\nu_{t}^{2}-\nu^{2}+\iu\Gamma_{0}\nu}\,.
\end{equation}
In the time domain this leads to the convolution \([n^2*E](t)\);
cf.~\eqref{eqwave2}. To avoid the expensive evaluation of the convolution
in~\eqref{eqwave2}, we derive an ADE. To this end we introduce the auxiliary
variable $P$, which is defined as
\begin{align}\label{eqadefreq}
  \frac{(n_{\Omega}^{2}-n_{\omega}^{2})
    \nu_{t}^{2}}{\nu_{t}^{2}-\nu^{2}+\iu\Gamma_{0}\nu}E &\eqqcolon P\,.\\
  \intertext{By this definition we obtain ${n(\nu)}^{2}E= n_{\omega}^{2}E + P$
    in the frequency domain. The second term of~\eqref{eqwave2} in the frequency
    domain becomes}
  \partial_{tt}({{n(t)}^{2} * E})=-\nu^2n_{\omega}E-\nu^2P\,.\label{eqsecterm}
  \intertext{
    From~\eqref{eqadefreq} we derive the ADE for the variable $P$ in the frequency domain}
  \nu_{t}^{2}P-\nu^{2}P+\iu\Gamma_{0}\nu P -
  (n_{\Omega}^{2}-n_{\omega}^{2})\nu_{t}^{2}E&=0\,.\label{eqadefreq2}
\end{align}
We transform~\eqref{eqadefreq2} to the time domain with an inverse Fourier
transform. This leads to the actual ADE~\eqref{eqwave:lor-nade2a} for Lorentz
materials in the time domain. We use the ADE~\eqref{eqadefreq2} to substitute
$-\nu^2P$ in~\eqref{eqsecterm} and insert this into the electromagnetic wave
equation~\eqref{eqwave2}, such that we recover~\eqref{eqwave2a} a
\begin{subequations}\label{eqwave:lor-nade2}
  \begin{align}
    \begin{split}\label{eqwave:lor-nade2a}
      \partial_{tt}P + \Gamma_{0}\partial_{t} P + \nu_{t}^2P -
      (n_{\Omega}^{2}-n_{\omega}^{2})\nu_{t}^2 E&=0\,,
    \end{split}
    \\
    \begin{split}\label{eqwave:lor-nade2b}
      -\Delta E + n_{\omega}^{2} \partial_{tt} E + (n_{\Omega}^{2}-
      n^{2}_{\omega})\nu_{t}^{2}E- \nu_{t}^{2}P -\Gamma_{0}\partial_{t}P+
      \chi^{(2)}\partial_{tt}(E^2)&=0\,,
    \end{split}
  \end{align}
\end{subequations}
with boundary and initial conditions as given in~\eqref{eqwave2b}. The technique
of avoiding the evaluation of the convolution by deriving an ADE is well-suited
for the space time finite element framework. However, alternative methods for
evaluating the convolution have been proposed, most notably the recursive
convolution method
(RCM)~\cite{dragnaGeneralizedRecursiveConvolution2015,liSemianalyticalRecursiveConvolution2020,jiaoTimedomainFiniteelementModeling2001}
and $z$-transform
method~\cite{abrahamDispersiveMobiusTransform2016,akbarzadeh-sharbafStableEfficientDirect2015}.

In~\cite{abrahamConvolutionFreeMixedFiniteElement2019} the $z$-transform
  has been applied to nonlinear dispersive media. This has proven to be
  efficient and easily generalizable to complex media, which can be hard to
  achieve when the ADE method is used. Nevertheless, the ADE method has
advantages over the RCM and $z$-transform approach. Firstly, the ADE will be
discretized and solved with the same numerical methods as the actual wave
equation. Thus we maintain the same order of approximation for all equations of
our model. Furthermore we maintain the high level of flexibility w.\,r.\,t.\ the
discretization techniques. This also facilitates error and stability
investigations. Secondly, the implementation effort for adding an ADE to an
existing space time finite element framework is small.

\subsection{Domain truncation}\label{sec:domain-truncation}
In numerical simuations, wave propagation and other physical studies have to be
truncated to bounded regions. Artificial boundaries that do not perturb the wave
propagation and lead to unphysical reflections, caused by the imposed boundary
conditions, are desired. Two main types of methods exist for the implementation
of such types of boundaries.

The first method uses an absorbing boundary condition on the surface of the
domain. The waves leave the domain without spurious reflections. In practice
these types of boundary conditions are difficult to implement due to their fast
growing complexity for higher order methods. Secondly, satisfactory absorption
at sharp incident angles independent of the order is not feasible.

The second method used for absorbing boundaries is the PML. It is based on the
idea to absorb all waves in an artificial boundary region using a volume
surrounding the physical domain. Here we apply the complex frequency shifted PML
(CFS-PML) that is based on~\cite{gedneyAuxiliaryDifferentialEquation2010}. It
allows us to derive ADEs similar to the Lorentz dispersion, which leads to a
high level of flexibility w.\,r.\,t.\ the applications of discretization
techniques.

\subsubsection{Complex frequency shifted PML}\label{sec:cfs-pml}
Complex frequency shifted PML (CFS-PML) are based on a complex stretch of
coordinates. As the name suggests, CFS-PML use a frequency-shifted complex
coordinate stretching function:
\begin{equation}\label{eq:cfspml1}
  s_{k}(\nu)
  =\kappa_{k} + \frac{\sigma_{k}}{\alpha_{k}+\iu\nu},\quad k\in {x,\,y,\,z}\,.
\end{equation}
The reciprocal of this function will be useful later on:
\begin{equation}\label{eq:cfspml2}
  \frac{1}{s_{k}(\nu)}
  =\frac{1}{\kappa_{k}} +
  \frac{\tilde{\sigma}_{k}}{\tilde{\alpha}_{k}+\iu\nu}
  ,\quad k\in {x,\,y,\,z},\text{ where }\; \tilde{\sigma_{k}}= -\frac{\sigma_{k}}{\kappa_{k}^{2}}
  \;\text{ and }\; \tilde{\alpha_{k}}=
  \frac{\sigma_{k}}{\kappa_{k}}+\alpha_{k}\,.
\end{equation}
In general, the parameters are functions depending on the position in the PML\@.
Here we chose a cosine profile
\begin{equation}
  f_{c}=b_{c} + a_{c}
  \paran{\frac{1}{2}-\frac{1}{2}\cos\paran{2\pi\frac{\vct{v}_{k}}{\delta}}}^{p_{c}},
  \quad\text{where}\quad c\in \{\alpha_{k},\,\sigma_{k},\,\alpha_{k}\}\,.
\end{equation}
The parameters should be chosen such that \(\kappa_{k}=1\) and \(\sigma_{k}=0\)
at the interface to the physical domain to ensure continuity of the physical
parameters and thereby the continuity of the solutions $E$ and $A$. The
frequency shift \(\alpha\) decays from the interface towards the PML boundary.
The parameter \(\delta\) is the thickness of PML and \(\vct{v}_{k}\) the
distance to the interface to the physical domain.

We derive the ADE formulation of the CFS-PML by multiplying the first term
of~\eqref{eqwave:lor-nade2b} by \(s_{x}\) which leads to
\begin{equation}\label{eqpml:lor-freqdomain1}
  \partial_{x}\frac{1}{s_{x}}\partial_{x} E + \paran{\kappa_{x} + \frac{\sigma_{x}}{\alpha_{x}+\iu\nu}}\paran{n^{2}_{\omega}
    + \frac{(n_{\Omega}^{2}-n_{\omega}^{2}) \nu_{t}^{2}}{\nu_{t}^{2}-\nu^{2}+\iu\Gamma_{0}\nu}}\nu^{2}E = 0\,.
\end{equation}
We need to derive an ADE for the term \(\frac{1}{s_{x}}\partial_{x} E\)
in~\eqref{eqpml:lor-freqdomain1} as well as for the product of \(s_{x}\) and the
Lorentz term. For the former we introduce the auxiliary variable \(Q\),
\begin{equation}\label{eqpml:lor-freqdomain2}
  \partial_{x}\frac{1}{s_{x}}\partial_{x} E = \partial_{x}\kappa_{x}^{-1}\partial_{x}E
  + \partial_{x}\frac{\tilde{\sigma}_{x}}{\tilde{\alpha}_{x}+\iu\nu}\partial_{x} E=
  \partial_{x}\kappa_{x}^{-1}\partial_{x} E + \partial_{x}Q,
\end{equation}
where Q is the solution to the equation
\(\iu\nu Q + \tilde{\alpha}_{x} Q - \tilde{\sigma}_{x} \partial_{x} E = 0\,.\)
The product of the Lorentz dispersion term is more involved. The product of PML
and Lorentz term yields
\begin{multline}
  \label{eqpml:lor-freqdomain1-2}
  \paran{\kappa_{x} + \frac{\sigma_{x}}{\alpha_{x}+\iu\nu}}\paran{n^{2}_{\omega}
    + \frac{(n_{\Omega}^{2}-n_{\omega}^{2}) \nu_{t}^{2}}{\nu_{t}^{2}-\nu^{2}+\iu\Gamma_{0}\nu}}=\\
  \paran{\kappa_{x} + \frac{\sigma_{x}}{\alpha_{x}+\iu\nu}}n^{2}_{\omega} +
  \kappa_{x}\frac{(n_{\Omega}^{2}-n_{\omega}^{2})
    \nu_{t}^{2}}{\nu_{t}^{2}-\nu^{2}+\iu\Gamma_{0}\nu}+
  \frac{\sigma_{x}}{\alpha_{x}+\iu\nu}\frac{(n_{\Omega}^{2}-n_{\omega}^{2})
    \nu_{t}^{2}}{\nu_{t}^{2}-\nu^{2}+\iu\Gamma_{0}\nu}\,.
\end{multline}
In the following definition of the equations we do not consider the last term
in~\eqref{eqpml:lor-freqdomain1-2}. This avoids the necessity to handle third
order time derivatives. Since the parameter \(\sigma\) will always be \(0\) at
the interface between physical and PML region, we don't expect this to have
major influence on our model and the accuracy of the simulations. We further
limit our model to the one dimensional case.

\begin{definition}{ADE-CFS-PML equation in 1 dimension with Lorentz
    dispersion}{lor-ade-cfs-pml}
  Let $\mathcal{D}=[0,\,L_{\mathcal{D}}]\subset \R$ be a closed interval with
  boundary \(\Gamma_{D}=\{0,\,L_{\mathcal{D}}\}\) and $I=(0,\,T]$ a bounded time
  interval. Corresponding to our real test case we prescribe Dirichlet boundary
  conditions on both endpoints of \(\mathcal{D}\). We further define the
  decomposition of \(\mathcal{D}\) into disjoint intervals
  \(\mathcal{D}=\mathcal{D}_{\text{Phy}}\cup\mathcal{D}_{\text{PML}}\). The
  decomposition is defined by the intervals
  \(\mathcal{D}_{\text{Phy}}=[0,\,L_{\text{Phy}})\) and
  \(\mathcal{D}_{\text{PML}}=[L_{\text{Phy}},\,L_{\mathcal{D}}]\), where
  \(L_{\text{PML}}=L_{\mathcal{D}}-L_{\text{Phy}}\). The physical region
  \(\mathcal{D}_{\text{Phy}}\) is the domain of physical interest, e.\,g.\, the
  periodically poled crystal in our case. The PML region
  \(\mathcal{D}_{\text{PML}}\) on the other hand purely exists for damping
  incoming waves. In the physical region $\Omega_{\text{Phy}}$ the governing
  equations are
  \begin{subequations}\label{eqpml:lor-wavesystem-phys}
    \begin{align}
      \partial_{tt}P + \Gamma_{0}\partial_{t} P
      + \nu_{t}^2P - (n_{\Omega}^{2}-n_{\omega}^{2})\nu_{t}^2 E&=0\,,\\
      -\Delta E + n_{\omega}^{2} \partial_{tt} E + (n_{\Omega}^{2}-
      n^{2}_{\omega})\nu_{t}^{2}E- \nu_{t}^{2}P -\Gamma_{0}\partial_{t}P+
      \chi^{(2)}\partial_{tt}(E^2)&=0\,,
    \end{align}
  \end{subequations}
  with initial conditions
  \(E(x,\,0) = E_{0}(x),\: x\in \mathcal{D}_{\text{Phy}}\) and boundary
  conditions \(E(x,\,t) = g(x,\,t),\: x=0,\: t\in (0,\,T]\). Inside the
  PML-region $\Omega_{\text{PML}}$ we solve the equations
  \begin{subequations}\label{eqpml:lor-ade-cfs-pml}
    \begin{align}
      \partial_{tt}P + \Gamma_{0}\partial_{t} P
      + \nu_{t}^2P - \kappa_{x}(n_{\Omega}^{2}-n_{\omega}^{2})\nu_{t}^2 E&=0\,,\\
      \partial_{t}R+\alpha_{x}R -n_{\omega}^{2}\sigma_{x}E&=0\,,\\
      \partial_{t} Q + \tilde{\alpha}_{x} Q - \tilde{\sigma}_{x} \partial_{x}
      E &= 0\,,\\
      -\nabla\cdot \kappa_{x}^{-1}\nabla E + \partial_{x} Q +
      \kappa_{x}n_{\omega}^{2} \partial_{tt} E +
      \kappa_{x}(n_{\Omega}^{2}- n^{2}_{\omega})\nu_{t}^{2}E&\\
      - \nu_{t}^{2}P -\Gamma_{0}\partial_{t}P +\partial_{t}
      \paran{n_{\omega}^{2}\sigma_{x}E - \alpha_{x}R}&=0\,,
    \end{align}
    with initial conditions \(E(x,\,0) = 0,\: x\in \mathcal{D}_{\text{PML}}\)
    and boundary conditions
    \(E(x,\,t) = 0,\: x=L_{\mathcal{D}},\: t\in (0,\,T]\).
  \end{subequations}
\end{definition}

\section{Discretization}\label{sec:disc}
In this section we develop the numerical methods for solving the electromagnetic
wave equation introduced in Def.~\ref{def:lor-ade-cfs-pml}. First we write the
electromagnetic wave equation as a first order in time system.
\begin{problem}{First order ADE-CFS-PML equation}{lor-ade-cfs-pml-fo}
  Considering Eq.~\eqref{eqpml:lor-wavesystem-phys} given in
  Def.~\ref{def:lor-ade-cfs-pml} the first order in time governing equations in
  the physical region $\mathcal{D}_{\text{Phy}}$ are
  \begin{subequations}
    \label{eqpml:lor-wavephys-first-order}
    \begin{align}
      \partial_{t}P + \Gamma_{0} P - U &=0\,,\\
      \nu_{t}^2P - (n_{\Omega}^{2}-n_{\omega}^{2})\nu_{t}^2 E+ \partial_{t}U&=0\,,\\
      n_{\omega}^{2} \partial_{t} E - \Gamma_{0} P+
      \chi^{(2)}\partial_{t}(E^2) - A&=0\,,\\
      -\Delta E + (n_{\Omega}^{2}- n^{2}_{\omega})\nu_{t}^{2}E-
      \nu_{t}^{2}P+\partial_{t}A &=0\,.
    \end{align}
  \end{subequations}
  Note that we introduced auxiliary variables $A$ and $U$ in order to obtain the
  first order formulation. The equations inside the PML-region
  $\Omega_{\text{PML}}$ can be formulated analogously without changing the
  equations for \(Q\) and \(R\).
\end{problem}

\subsection{Variatonal Space Time Methods}\label{sec:var-space-time-methods}
We briefly review variational space-time methods along the lines
of~\cite{kocherVariationalSpaceTime2014} and investigate discretization in space
and time with special focus on the nonlinearity and PML. In addition to the
definitions given in Def.~\ref{def:lor-ade-cfs-pml} let
$L\coloneqq L^2(\mathcal{D}) \;\text{and}\;V\coloneqq H^1_0(\Gamma_D;
\mathcal{D})$, where \(H_{0}^{1}\) is the space of \(H^{1}\)-functions with
vanishing trace on the specified part of the boundary (here \(\Gamma_{D}\)). For
the definition of these function spaces we refer
to~\cite{evansPartialDifferentialEquations2010}. We denote the \(L^{2}\)-inner
product by \(\dd{\bullet}{\bullet}_{\bullet}\), where the subscript denotes the
domain of integration and
\(\dd{\bullet}{\bullet}\coloneqq\dd{\bullet}{\bullet}_{\mathcal{D}}\). By
\(\norm{\bullet}\) we denote the \(L^{2}\)-norm and by \(\norm{\bullet}_{1}\)
the \(H^{1}\)-norm. Let \(L\coloneqq L^{2}(\mathcal{D})\) and
\(V\coloneqq H_{0}^{1}(\mathcal{D})\). We split the time interval $I$ into a
sequence of $N$ disjoint subintervals $I_n=(t_{n-1},\,t_n]$, $n=1,\dots,\,N$.
For a Banach space $B$ and $k\in \mathbb{N}_{0}$ we define
\begin{equation}
  \label{}
  \mathbb{P}_{k}(I_{n},\,B)=\brc{w_{\tau_{n}}\colon I_{n}\to B\suchthat
    w_{\tau_{n}}(t)=\sum_{j=0}^kW^jt^j\forall t\in I_n,\:W^j\in B \forall j}\,.
\end{equation}
For $r\in \mathbb{N}$ we define the finite element space that is built on the
spatial mesh as
\begin{equation}
  \label{eq:fespace}
  \begin{alignedat}{1}
    \mathcal{V}_{h}^{r}&=\brc{v_h\in C(\bar{\mathcal{D}})\suchthat v_h\vert_K \in
                         \mathcal{Q}_r(K)\;\forall K \in \mathcal{T}_h}\cap H_{0}^{1}(\mathcal{D}),\\
  \end{alignedat}
\end{equation}
where $\mathcal{Q}_r(K)$ is the space defined by the reference mapping of
polynomials on the reference element with maximum degree $r$ in each variable.
For the discretization of the equations
in~\eqref{eqpml:lor-wavephys-first-order} we split the time interval \(I\) into
a sequence of \(N\) subintervals \(I_{n}\). With a discontinuous test basis
supported on the subintervals $I_n$, this leads to a sequence of the following
local problems on each subinterval.
\begin{problem}{Space-time formulation of the ADE-CFS-PML
    equation}{lor-ade-cfs-stm}
  Assume that the trajectories
  $e_{\tau,\,h},\:a_{\tau,\,h},\:p_{\tau,\,h}$ and $u_{\tau,\,h}$ have been
  computed before for all $t \in [0,\,t_{n-1}]$, starting with initial
  conditions
  $e_{\tau,\,h}(0)=e_{0,\,h},\:a_{\tau,\,h}(0)=a_{0,\,h},\:p_{\tau,\,h}(0)=p_{0,\,h}$
  and $u_{\tau,\,h}(0)=u_{0,\,h}$. Consider solving the following local
  problem on the interval $I_n$:
  
  For given
  \(e_{h}^{n-1} \coloneqq e_{\tau,\,h}(t_{n-1})\in \mathcal{V}_{h}^{1}\) with
  \(e_{\tau,\,h}(t_{0})\coloneqq e_{0,\,h}\), find
  \((e_{\tau,\,h},\,a_{\tau,\,h},\,p_{\tau,\,h},\,u_{\tau,\,h}) \in
  \mathbb{P}_{1}(I_{n},\,\mathcal{V}_{h}^{1})\times{\mathbb{P}_{1}(I_{n},\,\mathcal{V}_{h}^{1})}^{3}\),
  such that \(e_{\tau,\,h}(t_{n-1})=e_{h}^{n-1}\) and
  \begin{subequations}\label{eq:wavephy-variational}
    \begin{align}
      \int_{t_{n-1}}^{t_{n}}\dd{\partial_{t}p_{\tau,\,h}}{\varphi_{\tau,\,h}} + \Gamma_{0} \dd{p_{\tau,\,h}}{\varphi_{\tau,\,h}} - \dd{u_{\tau,\,h}}{\varphi_{\tau,\,h}} \drv t&=0\,,\label{eq:wavephy-variational-a}\\
      \int_{t_{n-1}}^{t_{n}}\nu_{t}^2\dd{p_{\tau,\,h}}{\varphi_{\tau,\,h}} -
      (n_{\Omega}^{2}-n_{\omega}^{2})\nu_{t}^2\dd{e_{\tau,\,h}}{\varphi_{\tau,\,h}} + \dd{\partial_{t}u_{\tau,\,h}}{\varphi_{\tau,\,h}}\drv t&=0\,,\label{eq:wavephy-variational-b}\\
      \notag\int_{t_{n-1}}^{t_{n}}n_{\omega}^{2} \dd{\partial_{t} e_{\tau,\,h}}{\varphi_{\tau,\,h}} - \Gamma_{0}
      \dd{p_{\tau,\,h}}{{\varphi_{\tau,\,h}}}\qquad\qquad&\\
      +\chi^{(2)}\dd{\partial_{t}(e_{\tau,\,h}e_{\tau,\,h})}{\varphi_{\tau,\,h}} - \dd{a_{\tau,\,h}}{\varphi_{\tau,\,h}}\drv t&=0\,,\label{eq:wavephy-variational-c}\\
      \notag\int_{t_{n-1}}^{t_{n}}\dd{\nabla e_{\tau,\,h}}{\nabla\varphi_{\tau,\,h}} + (n_{\Omega}^{2}-
      n^{2}_{\omega})\nu_{t}^{2}\dd{e_{\tau,\,h}}{\varphi_{\tau,\,h}}\qquad\qquad&\\
      -
      \nu_{t}^{2}\dd{p_{\tau,\,h}}{\varphi_{\tau,\,h}}+\dd{\partial_{t}a_{\tau,\,h}}{\varphi_{\tau,\,h}}\drv t&=\int_{t_{n-1}}^{t_{n}}f\,.\label{eq:wavephy-variational-d}
    \end{align}
  \end{subequations}
  for all
  \((v_{\tau,\,h},\,w_{\tau,\,h},\,w_{\tau,\,h},\,w_{\tau,\,h}) \in
  \mathbb{P}_{0}(I_{n},\,\mathcal{V}_{h}^{1})\times{\mathbb{P}_{0}(I_{n},\,\mathcal{V}_{h}^{1})}^{3}\).
  The equations inside the PML region can be derived in an analogous manner with
  \(R\in\mathbb{P}_{1}(I_{n},\,\mathcal{V}_{h}^{1})\) and
  \(Q\in\mathbb{P}_{1}(I_{n},\,\mathcal{V}_{h}^{1})\).
\end{problem}
We formulate a discrete problem in space and time by expanding the solutions
\(e_{\tau,\,h}\), \(a_{\tau,\,h}\), \(p_{\tau,\,h}\), \(u_{\tau,\,h}\)
in~\eqref{eq:wavephy-variational} in terms of temporal basis functions. In this
work we consider the lowest order case of linear polynomials. A detailed
derivation can be found in the Appendix~\ref{sec:app-fully-disc}.

\subsubsection{A linear Galerkin method in time for the wave equation}\label{sec:linear-galerkin-method}
The representation of
\((e_{\tau,\,h},\,a_{\tau,\,h},\,p_{\tau,\,h},\,u_{\tau,\,h})\) in terms of
linear polynomials \(\varphi_{n,0}\) and
\(\varphi_{1,1}\in \mathbb{P}_{1}(\bar{I}_{n};\,\R)\) on \(\bar{I}_{n}\) can be
substituted into~\eqref{eq:wavephy-variational};
cf.~Appendix.~\ref{sec:app-fully-disc}. Then we
integrate~\eqref{eq:wavephy-variational} numerically and get the system of
equations
\begin{equation}
  \label{eqcg1}
  \mathcal{M}\vct{v}_{h}^{n}
  +\frac{\tau_{n}}{2}\mathcal{A}\vct{v}_{h}^{n}
  =\frac{\tau_{n}}{2} \paran{\vct{b}_{h}^{n-1} + \vct{b}_{h}^{n}}
  + \mathcal{M}\vct{v}_{h}^{n-1}
  -\frac{\tau_{n}}{2}\mathcal{A}\vct{v}_{h}^{n-1}\,.
\end{equation}
The solution vectors are defined as
\(\vct{v}_{h}^{n}=\paran{{(u_{h}^{n})}^{\top},{(p_{h}^{n})}^{\top},{(e_{h}^{n})}^{\top},{(a_{h}^{n})}^{\top}}^{\top}\).
The discrete operators \(\mathcal{M}\) and \(\mathcal{A}\) are defined in terms
of \({(\mat{M})}_{ij}=\dd{\varphi_{h,\,j}}{\varphi_{h,\,i}}\),
\({(\mat{A})}_{ij}=\dd{\nabla\varphi_{h,\,j}}{\nabla\varphi_{h,\,i}}\) and the
nonlinear operator
\({(\mat{N}(e))}_{ij}=\dd{e \varphi_{h,\,j}}{\varphi_{h,\,i}}\) as follows
\begin{equation*}
  \mathcal{A}=\begin{pmatrix}
                -\mat{M} & \Gamma_{0}\mat{M} & 0 & 0\\
                0 &\nu_{t}^{2}\mat{M} & -(n_{\Omega}^{2}-n_{\omega}^{2})\nu_{t}^{2}\mat{M} &
                                                                                             0\\
                0 & \Gamma_{0}\mat{M} & 0 & -\mat{M}\\
                0 &-\nu_{t}^{2}\mat{M} &
                                         (n_{\Omega}^{2}-n_{\omega}^{2})\nu_{t}^{2}\mat{M}+\mat{A}
                                                 & 0\\%
              \end{pmatrix}\,,\; %
              \mathcal{M}=\begin{pmatrix}
                            0 & \mat{M} & 0 & 0\\
                            \mat{M} & 0 & 0 & 0\\
                            0 & 0 & n_{\omega}^{2}\mat{M}+\chi^{(2)}\mat{N}(e) &0\\
                            0 & 0 & 0 & \mat{M}
                          \end{pmatrix}\,.
\end{equation*}
The system of equations~\eqref{eqcg1} is equivalent to the one we would get from
the well-known Crank-Nicholson method. By deriving update equations for the
variables \(u_{h}^{n},\:p_{h}^{n}\) and \(a_{h}^{n}\) we can
condense~\eqref{eqcg1} such that we just need to solve for \(e_{h}^{n}\);
cf.~Appendix~\ref{sec:app-condense-sys}.

For the solution of the system of nonlinear equations~\eqref{eqcg1} we use a
linearization with a damped version of the Newton method. Let
\(\vct{v}_{h}^{n}\in \mathcal{V}_{h} \colon
\boldsymbol{\mathcal{S}}(\vct{v}_{h}^{n})=\mat{F}\) be the nonlinear system of
equations defined by~\eqref{eqcg1}. We assume that it is sufficiently
differentiable by means of the Gateaux derivative
\(\boldsymbol{\mathcal{S}}' (\vct{v}_{h}^{n};\,\delta \vct{v}_{h}^{n})\coloneqq
\frac{\drv}{\drv s} \boldsymbol{\mathcal{S}}(\vct{v}_{h}^{n}+\varepsilon \delta
\vct{v}_{h}^{n})\vert_{\varepsilon=0}\). \(\boldsymbol{\mathcal{S}}'\) denotes
the derivative of \(\boldsymbol{\mathcal{S}}\) at
\(\vct{v}_{h}^{n}\in \mathcal{V}_{h}\) in direction
\(\delta\vct{v}_{h}^{n}\in \mathcal{V}_{h}\). In each step of the Newton method
a linear system of equations arise from the discretization, which can then be
solved with linear solvers, e.\,g.\ Conjugate Gradient or Multigrid methods. For
a general overview of iterative methods
see~\cite{saadIterativeMethodsSparse1996} or
\cite{hackbuschMultiGridMethodsApplications1985,wesselingIntroductionMultigridMethods1991}
for multigrid methods in particular. Details on the application of Newton's
method and its variants to nonlinear partial differential equations can be found
in the
literature~\cite{deuflhardNewtonMethodsNonlinear2011,deuflhardNewtonMethodsNonlinear2005}.
Similar methods for the solution of nonlinear electromagnetic
wave equations including dispersive nonlinearities have been developed in~\cite{abrahamConvolutionFreeFiniteElementTimeDomain2019,abrahamConvolutionFreeMixedFiniteElement2019}.
They have also been implemented on GPUs~\cite{abrahamParallelFiniteElementTimeDomain2020}.
The Newton iteration for solving~\eqref{eqcg1} with an initial guess
\({\vct{v}_{h}^{n}}_{0}\in \mathcal{V}_{h}\) iterates for \(m=0,\dots\)
\begin{equation}\label{gcc-newton}
  \delta{\vct{v}_{h}^{n}}_{m}\colon
  \boldsymbol{\mathcal{S}}'({\vct{v}_{h}^{n}}_{m-1};\,\delta{\vct{v}_{h}^{n}}_{m}) =
  \mathbf{F}-\boldsymbol{\mathcal{S}}({\vct{v}_{h}^{n}}_{m-1}),\quad{\vct{v}_{h}^{n}}_{m}\coloneqq {\vct{v}_{h}^{n}}_{m-1} + \delta{\vct{v}_{h}^{n}}_{m}\,.
\end{equation}

As we have already seen in~\ref{eqpml:lor-ade-cfs-pml} the system of equations
grow larger in the PML region due to the additional variables for the PML.
However, we can condense this system to the same size as for the physical domain
by reducing the ADEs to simple update equations. Since we don't consider
nonlinear processes inside the PML, the nonlinear part can be left out here. We
redefine \(\vct{v}_{h}^{n}\) and the operators \(\mathcal{M}\) and
\(\mathcal{A}\) in the PML region \(\mathcal{D}_{\text{PML}}\) as
\(\vct{v}_{h}^{n}=\paran{{(q_{h}^{n})}^{\top},{(r_{h}^{n})}^{\top},
  {(u_{h}^{n})}^{\top},{(p_{h}^{n})}^{\top},{(e_{h}^{n})}^{\top},
  {(a_{h}^{n})}^{\top}}^{\top}\) and%
\begin{equation*}
  \scalemath{.8}{
    \mathcal{A}=
    \begin{pmatrix}
      \tilde{\alpha}G& 0&  0&  0&  \tilde{\sigma}\mat{A}&  0\\
      0&  \alpha\mat{M}&  0&  0&  -n_{\omega}^{2}\sigma\mat{M}& 0\\
      0&  0&  -\mat{M}& \Gamma_{0}\mat{M}&  0&  0\\
      0&  0&  0&  \nu_{t}^{2}\mat{M}& -(n_{\Omega}^{2}-n_{\omega}^{2}) \kappa \nu_{t}^{2}\mat{M}&  0\\
      0&  -\alpha\mat{M}& 0&  -\Gamma_{0}\mat{M}&   n_{\omega}^{2} \sigma\mat{M}&  -\mat{M}\\
      G& 0& 0& -\nu_{t}^{2}\mat{M}& (n_{\Omega}^{2}-n_{\omega}^{2}) \kappa
                                    \nu_{t}^{2}\mat{M}+\frac{1}{\kappa}\mat{A}& 0
    \end{pmatrix}\,,\:%
    \mathcal{M}=
    \begin{pmatrix}
      \mat{G}&  0&  0&  0&  0&  0\\
      0&  \mat{M}&  0&  0&  0&  0\\
      0&  0&  0&  \mat{M}&  0&  0\\
      0&  0&  \mat{M}&  0&  0&  0\\
      0&  0&  0&  0&  n_{\omega}^2\kappa\mat{M}&  0\\
      0& 0& 0& 0& 0& \mat{M}
    \end{pmatrix}\,. }
\end{equation*}
Using these definitions, the notation of the system of equations in the PML
region doesn't differ compared to~\eqref{eqcg1}. Analogous to the equations in
the physical region we can derive vector updates for the auxiliary variables.
The condensed system is also presented in the appendix. In our numerical
investigations in Sec.~\ref{sec:num-sim} we study the computational effort
required for the PML.

\section{Numerical simulation}\label{sec:num-sim}
In this section we demonstrate the efficiency of our techniques and the PML in
particular. Firstly, we verify the accuracy of our methods with convergence
tests. We chose a test based on plane waves\@. Secondly, we investigate the
physical problem of THz generation in PPLN\@. We verify the methods once again
by comparing results and quantities of interest to experimental data. We discuss
the practical implications, with a focus on aspects not yet investigated,
e.\,g.\ second harmonic generation.

For the implementation we use the finite element toolbox
deal.II~\cite{arndtDealIILibrary2021} along with the Trilinos library.
These
software libraries support parallelization with MPI which is used throughout
this work. The nonlinear systems of equations are solved
with a Newton-Krylov method. The linear systems of equations arising in each
Newton step are solved with the conjugate gradients (CG) method. The convergence is
accelerated by the algebraic multigrid solver MueLU~\cite{MueLu},
that is used as a preconditioner with a single sweep performed for every CG step.
\begin{figure}[ht!]
  \centering
  \begin{tikzpicture}[xscale=5.5,yscale=7,domain=-0.5:0,>=latex]
    \draw[color=mpblue, samples=2000]
    plot[smooth](\x,{.125+.1*exp(-((\x+0.25)*(\x+0.25))*100)*cos(300*\x r)}); %
    \node (a) at (-0.425,0.175) {$\omega$}; %
    \draw[color=mpblue, samples=2000, xshift=1.83cm] plot[smooth]
    (\x,{.225+.075*exp(-((\x+0.25)*(\x+0.25))*100)*cos(300*\x r)}); %
    \node (b) at (1.88,0.25) {$\omega$}; %
    \draw[color=mporange, samples=2000, xshift=1.83cm] plot[smooth]
    (\x,{.075+.05*cos(30*\x r)}); %
    \node[fill=white] (c) at (1.88,0.075) {$\Omega$}; %
    \fill [color=mpblue!15] (0,0) rectangle (0.2,0.3); %
    \fill [color=mpblue!7] (0.2,0) rectangle (0.4,0.3); %
    \fill [color=mpblue!15] (0.4,0) rectangle (0.6,0.3); %
    \fill [color=mpblue!7] (0.9,0) rectangle (1.1,0.3); %
    \fill [color=mpblue!3] (1.1,0) rectangle (1.33,0.3); %
    \draw[|<->|, decorate, decoration={amplitude=2pt}] (0,0.275) --
    node[fill=mpblue!10,pos=0.4,inner sep=0pt]{$\Lambda$} ++ (0.4,0.); %
    \path[draw=black] (0,0) -- ++(1.33,0) -- node[pos=0.5,
    right]{$\Gamma_{\text{out}}$} ++(0,0.3) -- ++(-1.33,0) -- node[pos=0.33,
    left,fill=white,inner sep=1pt]{$\Gamma_{\text{in}}$}cycle; %
    \draw[->] (.1,.025)-- ++ (0,0.225); %
    \draw[<-] (.3,.025) -- node[fill=mpblue!7,inner
    sep=1pt,pos=0.5]{$-\chi^{(2)}$}++ (0,0.225); %
    \draw[->] (.5,.025) -- node[fill=mpblue!15,inner
    sep=1pt,pos=0.5]{$\chi^{(2)}$} ++ (0,0.25); %
    \draw[<-] (1.0,.025) -- ++ (0,0.25); %
    \draw[dotted,line width=1pt] (0.7,.15) --++(.1,0); %
    \node (pml) at (1.2,0.15){PML}; %
  \end{tikzpicture}
  \caption{\label{fig:qpmexample}Example setting of a periodically poled crystal
    with period \(\Lambda\). The pump pulse $g(t)$ at frequencies
    \(\omega_{1,\,2}\) enters the crystal on the left side
    \(\Gamma_{\text{in}}\). In the subsequent layers THz radiation is
    generated.}
\end{figure}
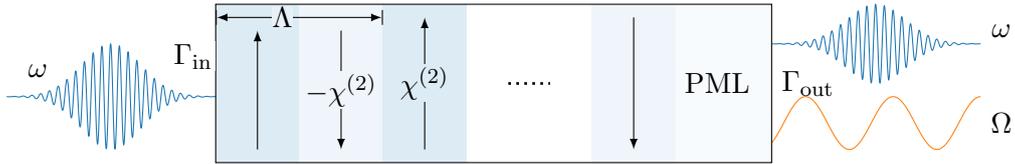

\subsection{Convergence and runtime study}\label{sec:conv-study}
Here we verify the numerical methods we developed before. To this end we
manufacture a solution by prescribing a function as the solution
to~\eqref{eqpml:lor-wavephys-first-order}. We use the residual of this function
as a source term, which in turn makes the prescribed function the solution. We
use the linear Galerkin method proposed in~\eqref{eqcg1} for the time
discretization and the finite element space $V_{h}^{2}$ defined
in~\eqref{eq:fespace} for the spatial discretization. Consequently, we expect
second-order convergence. Although piecewise linear polynomials in space would
have been sufficient for ensuring this rate of convergence, we use piecewise
quadratic polynomials which has been shown to be advantageous.
  
We chose a 1D test case in the domain \(\mathcal{D}=[0,\,\num{0.001955}]\) over
the time interval \(I=[0,\,\num{1e-13}]\). As the electric field we chose
\begin{equation}
  \label{eq:testefield1}
  E(x,\,t)=\sin \left(2 \pi \omega_{2} \left(x-n_{2} t\right)\right)+
  \sin \left(2 \pi \omega_{1} \left(x-n_{1} t\right)\right)\,.
\end{equation}
The source term can be calculated by substituting~\eqref{eq:testefield1}
into~\eqref{eqpml:lor-wavephys-first-order}. To compute the error in the
physical domain and exclude error contributions from within the PML region, we
introduce a weighting function \(l\colon \mathcal{D} \to \R\) that is 1 in the
physical domain and 0 in the PML region:
\[
  l(x)=
  \begin{cases} 0,\;x\in\mathcal{D}_{\text{PML}}\,,\\
    1,\;x\in\mathcal{D}_{\text{Phy}}\,.
  \end{cases}
\]
Furthermore we multiply \(l\) by the source term to restrict it to the physical
domain. Thereby the solution inside \(\mathcal{D}_{\text{Phy}}\) is given
by~\eqref{eq:testefield1}. Then it propagates into \(\mathcal{D}_{\text{PML}}\)
where it is attenuated to the point of vanishing.

\begin{figure}[ht!]
  \includegraphics[width=.95\linewidth]{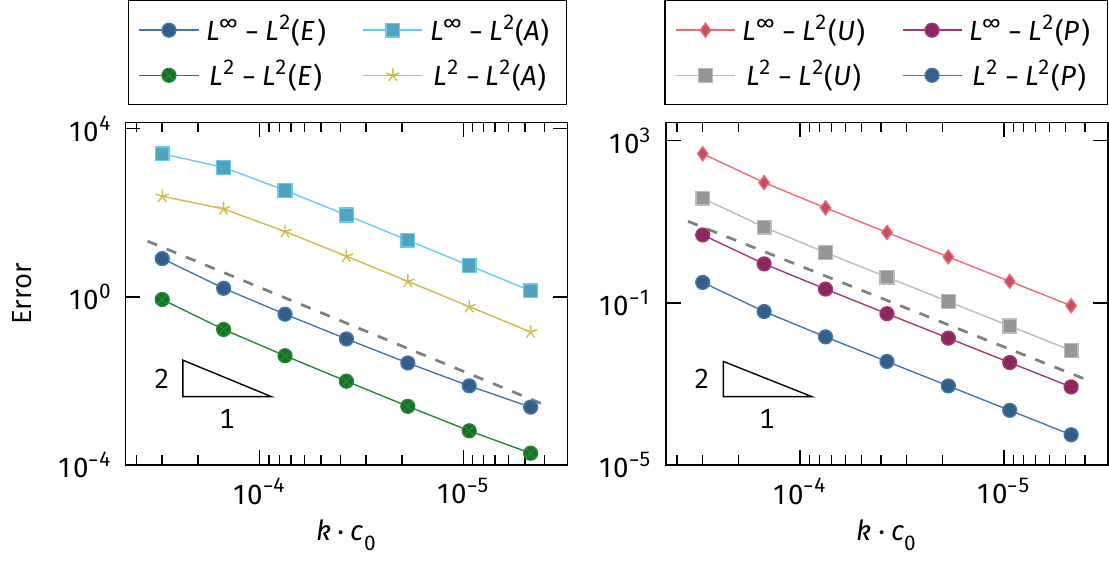}
  \caption{Calculated errors for the electric field $E$ and the auxiliary variables
    \(A,\:U,\:P\) for the linear Galerkin method plotted against the timestep-size. The expected quadratic order of
 convergence is visualized by the dashed line and the triangles in the lower
 left corners.\label{fig-condpml1}}  
\end{figure}

We study the errors $e_{Z}=Z(x,\,t)-Z_{\tau,\,h}(x,\,t)$ for $Z\in\{E,\,A,\,P,\,U\}$ in the norms
\begin{equation}
  \label{}
  \norm{e_Z}_{L^{\infty}(L^2)}=\max_{t\in I}\paran{ \int_{\mathcal{D}} \lvert e_{Z}\rvert^{2}\drv x}^{\frac{1}{2}}\,,\quad
  \norm{e_Z}_{L^2(L^2)}=\paran{\int_{I}\int_{\mathcal{D}} \lvert
    e_{Z}\rvert^{2}\drv x\drv t}^{\frac{1}{2}}\,.
\end{equation}
We abbreviate the error quantity
$\norm{e_E}_{L^{\infty}(L^2)}$ by $L^{\infty}-L^{2}(E)$ and analogously for the
other norms and variables. The errors are calculated by simultaneous
refinement in space and time. 
Table~\ref{tab-condpml1} and Fig.~\ref{fig-condpml1} show the optimal order of
convergence we expected, due to the second order convergence in time. This proves the efficiency of our PML
implementation and that it is suitable for nonlinear dispersive materials.
Towards the end we see a decrease in the order of convergence for the electric
field, which is due to minor reflections caused by the PML. This can be avoided
at the cost of increased computation time by extending the PML. Here we decided
to chose the PML similar to the physical problem we investigate later.
\begin{table}[htbp]
  \centering {\footnotesize
  \begin{tabular}{llrlrlrlr}
    \toprule
    \(k\cdot c_{0}\) & \(L^{\infty}-L^2 (E)\) &EOC  & \(L^{\infty}-L^2 (A)\) &EOC  & \(L^2-L^2 (E)\) & EOC & \(L^2-L^2 (A)\) &EOC \\
    \midrule
    \num{2.9979e-04} & \num{8.1623e+00} &  --- & \num{2.5475e+03} & --- & \num{8.6939e-01} & --- & \num{2.4871e+02} & ---\\
    \num{1.4990e-04} & \num{1.6054e+00} & 2.35 & \num{1.2074e+03} & 1.08 & \num{1.6802e-01} & 2.37 & \num{1.2409e+02} & 1.00\\
    \num{7.4948e-05} & \num{3.8809e-01} & 2.05 & \num{3.4400e+02} & 1.81 & \num{3.9945e-02} & 2.07 & \num{3.5988e+01} & 1.79\\
    \num{3.7474e-05} & \num{1.0013e-01} & 1.95 & \num{8.8535e+01} & 1.96 & \num{9.8922e-03} & 2.01 & \num{9.3145e+00} & 1.95\\
    \num{1.8737e-05} & \num{2.6876e-02} & 1.90 & \num{2.2321e+01} & 1.99 & \num{2.4962e-03} & 1.99 & \num{2.3478e+00} & 1.99\\
    \num{9.3685e-06} & \num{7.6661e-03} & 1.81 & \num{5.5903e+00} & 2.00 & \num{6.5436e-04} & 1.93 & \num{5.8804e-01} & 2.00\\
    \num{4.6843e-06} & \num{2.3916e-03} & 1.68 & \num{1.3984e+00} & 2.00 & \num{1.9160e-04} & 1.77 & \num{1.4706e-01} & 2.00\\
    \bottomrule
  \end{tabular}}
  \caption{Calculated errors and experimental orders of convergence (EOC) for the electric field and the auxiliary variable
    \(A\) for the linear Galerkin method introduced in
    Sec.~\ref{sec:linear-galerkin-method}.\label{tab-condpml1}}  
\end{table}
We use the convergence test to investigate the influence of the PML on the
runtimes and demonstrate the efficiency of the methods used. To this end we
compare the computation times per degree of freedom (DoF) inside the PML with
those in the physical domain. In order to minimize the computational overhead of
the PML we assemble the data structures for the PML once and reuse them in every
timestep. This is possible since the PML does not have any nonlinearity which
needs to be evaluated. For example, the evaluation of the PML damping functions
$\alpha$, $\sigma$, $\kappa$ can become expensive. However, they are time
independent and only need to be evaluated once. By caching them the time spent
per DoF in the PML increases by less than \(1\%\) compared to a DoF in the
physical region\@.

This shows that the PML is resource efficient, scales well and the computational
overhead is solely determined by the size of the PML\@. In the following section
we briefly analyze the cost of the PML in a realistic setting.
\subsection{THz generation in PPLN}\label{sec:THz-generation-in-PPLN}
\begin{figure}[!ht]
  \includegraphics[width=\linewidth]{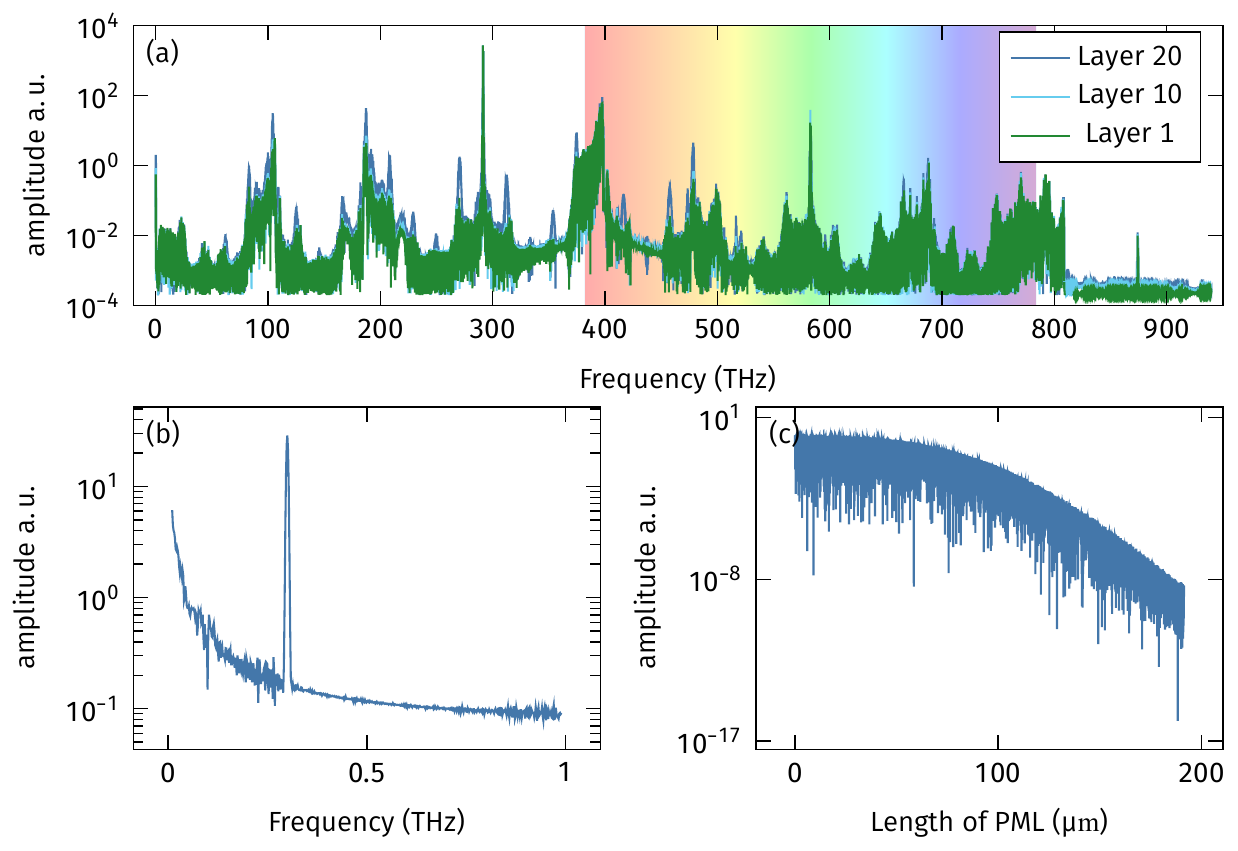}
  \caption{Plot (a) shows the spectrum of the electromagnetic field after it has passed 25
    periods of the crystal shown in Fig.~\ref{fig:qpmexample} in the
    $0\,\si{\tera\hertz}$ to $\SI{1000}{\tera\hertz}$ region and (b) in the
    region from $\SI{0.01}{\tera\hertz}$ to $\SI{100}{\tera\hertz}$. Subplot (c)
    shows the intensity of the electric field inside the PML at time
    \(t=\SI{750}{\femto\second}\).}\label{fig:freqplot1}
\end{figure}
\begin{figure}[!ht]
  \includegraphics[width=\linewidth]{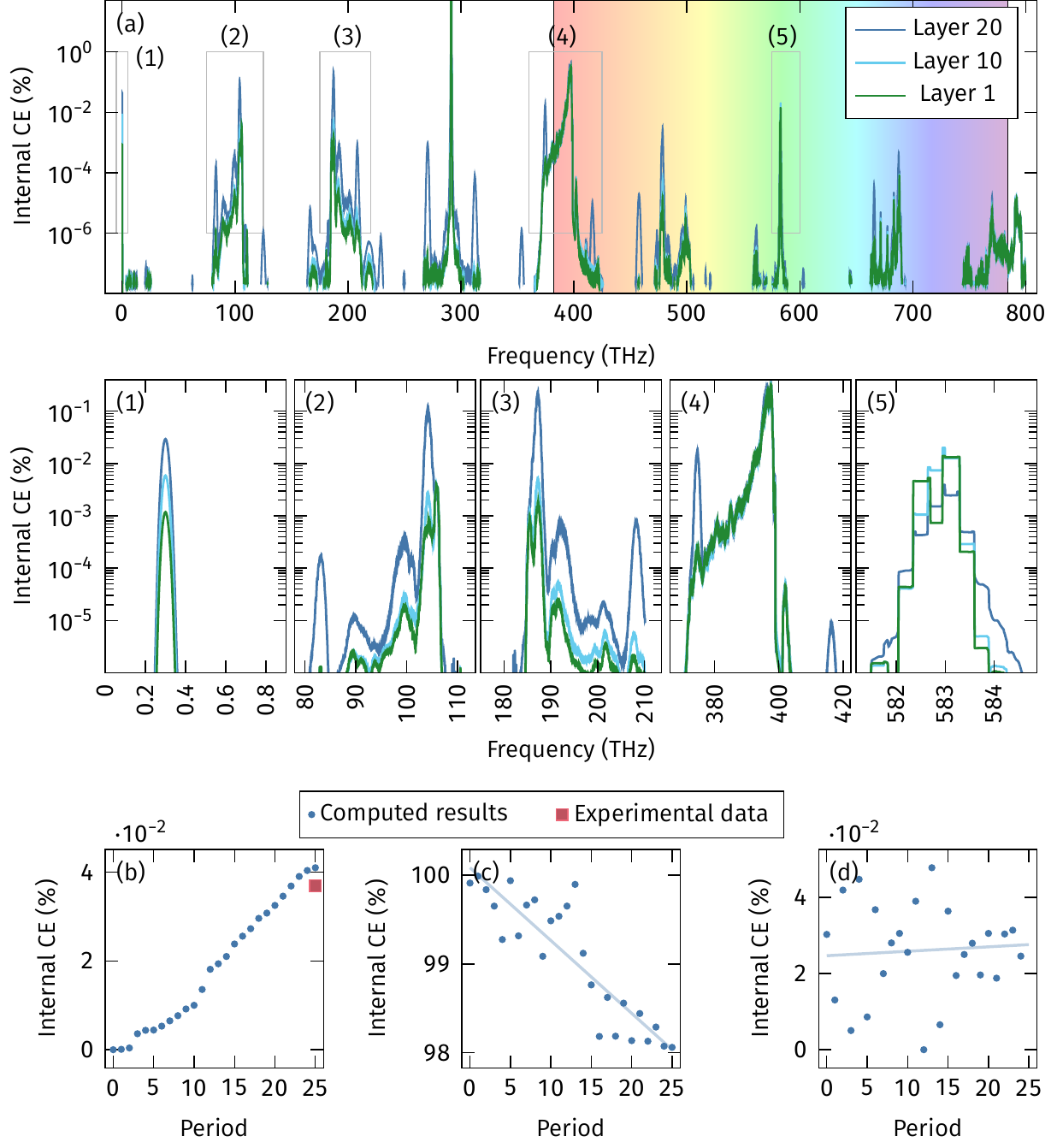}
  \caption{Internal conversion efficiency (a) for the frequencies from
    $\SI{100}{\tera\hertz}$ to $\SI{800}{\tera\hertz}$. In subplot (b) the
    intensity at frequencies between $\SI{0.1}{\tera\hertz}$ to
    $\SI{1}{\tera\hertz}$ are summed up, then the corresponding conversion
    efficiencies are plotted over the 25 layers. Subplot (c) shows the
    conversion efficiencies determined in the same way for frequencies between
    $\SI{100}{\tera\hertz}$ and $\SI{800}{\tera\hertz}$. Subplot (d) shows the
    conversion efficiencies for the second harmonic. In subplots (c) and (d) the
    pale blue lines show the linear regression of the
    data.}\label{fig:freqplot2CE}
\end{figure}
In our numerical investigations we simulate THz generation in PPLN\@. The main
goal in this section is to reproduce the experimental results presented
in~\cite{olgunHighlyEfficientGeneration2022}. The simulation data has the
advantage that we can evaluate it in all points of the space-time domain. Thus
we can correlate the different harmonics with each other and uncover effects of
the periodic poling and quasi phase matching. This will improve the
understanding of the physical processes. That potentially leads to the
improvement of the experimental setup especially for higher intensities where
simplified models fail. As in~\cite{olgunHighlyEfficientGeneration2022}, we use
two pump pulses with super Gaussian envelope
\[g(t)=\exp\big(-{\big(2\log
    2{\big(\tfrac{t}{\tau}\big)}^{2}\big)}^{P}\big)(\cos(2\pi\omega_1 t)+
  \cos(2\pi \omega_2 t))\,.\] The pulses are separated in center frequency by
the THz frequency with a full width half maximum \(\tau=\SI{250}{\nano\second}\)
and frequencies \(\omega_{1}=\SI{291.56}{\tera\hertz}\),
\(\omega_{2}=\SI{291.26}{\tera\hertz}\) and \(P=6\). We chose a pulse with
average fluence of \(\SI{2000}{\joule\per\metre}\). It is applied at the
left-hand side of the crystal by a Dirichlet boundary condition and then
propagates through the domain until the PML is hit on the right-hand side. The
problem setting is sketched in Fig.~\ref{fig:qpmexample}. The computational
effort for these simulations are very high: Depending on the fluence, which
influences the number of iterations of the newton method, the simulations can
take up to 4 weeks. The simulations presented here took 3 weeks on a workstation
with 2 Intel Xeon E5-2699 CPUs ($2\times18$ cores). In this study, we limited
our investigations and numerical simulations to one spatial dimension. This was
necessitated by simulation times and the added complexity of using Perfectly
Matched Layers (PMLs) in 2D and 3D. In the settings investigated here, the
simplification of reducing the simulations to one spatial dimension and
neglecting the impacts of the remaining spatial directions is not expected to
significantly perturb the results. The simulation results presented here are
based on a timestep size of $k=\num{5.e-17}$ and average cell-size of
$\num{5.175e-8}$, which leads to $\num{1.e10}$ number of timesteps and
$\num{425985}$ degrees of freedom in space.

The intensities of the simulated results
are presented in Fig.~\ref{fig:freqplot1}. We observe, besides the THz radiation
generated by difference frequency generation (DFG) in
Fig.~\ref{fig:freqplot1}~(b), the harmonics in the optical domain in
Fig.~\ref{fig:freqplot1}~(a). The optical harmonics are simultaneously generated
by SHG and sum frequency generation (SFG).
In particular the second harmonic near
\(\SI{600}{\tera\hertz}\) and the other \(n\)th harmonics can be well observed.
At \(\SI{900}{\tera\hertz}\) we can observe the effect of DFG from the
harmonics. The ability to simulate this phenomenon with such a low conversion
efficiency again demonstrates the potential and accuracy of the proposed method.
We also find significant intensities at frequencies
\SIlist{100;200;400}{\tera\hertz}; cf.~\ref{fig:freqplot2CE}~(a),
\ref{fig:freqplot1}~(a).

In Fig.~\ref{fig:freqplot1}~(c) we show the intensity over the length of the PML at
time \(t=\SI{250}{\pico\second}\) when the pulse has entered the crystal. The
intensity decays monotonically over the length of the crystal until it almost
vanishes. Reflections from the boundary are damped again. Therefore, based on
Fig.~\ref{fig:freqplot1}~(c) we don't expect spurious reflections into the
physical domain. Meanwhile the solution stays stable, which proves the
effectiveness of the formulation of the PML\@. The PML has the same length as
one poling period of the PPLN crystal. With a period of
\(\Lambda=\SI{212}{\micro\meter}\) and 25 periods the computational domain is
enlarged by about 4\%. Considering that the increase in computational cost per
degree of freedom was about 1\% inside the PML in the convergence test, the PML
accounts for 4\% of the total computing time, which is reasonable.

In Fig.~\ref{fig:freqplot2CE}~(a) we show the internal conversion efficiencies
plotted over the whole spectrum from \SIrange{0}{800}{\tera\hertz}. The
methods and physical models we use allow us to include nonlinear effects apart
from optical to THz conversion in our results. Near \(\SI{0.3}{\tera\hertz}\) we
see the remaining parts of the two input pulses. We observe a narrow peak near
$0\si{\tera\hertz}$, corresponding to the generated THz radiation at
\(\SI{0.3}{\tera\hertz}\). We further see the generation of optical harmonics in
a cascading manner. These efficiencies are mostly constant, since
these processes are phase-mismatched. For example, the SHG is phase
mismatched and nevertheless the conversion efficiencies are of similar magnitude
compared to the ones at \(\SI{0.3}{\tera\hertz}\). This is due to the much
higher efficiency for SHG than for $\si{\tera\hertz}$ generation and despite the
large phase mismatch some is generated.

Similar effects can be observed in Fig.~\ref{fig:freqplot2CE}~(a) where the conversion
efficiency in the layers are plotted. The conversion efficiency at
$\SI{400}{\tera\hertz}$ remains constant over the layers; cf.~Fig.~\ref{fig:freqplot2CE}~(4). The
\(\SI{0.3}{\tera\hertz}\) radiation on the other hand grows over the layers since
the process is phase-matched;
cf.~Fig.~\ref{fig:freqplot2CE}~(1),~\ref{fig:freqplot2CE}~(b).
In~Fig.~\ref{fig:freqplot2CE}~(2,3) we observe similar phenomena at
approximately $\SI{100}{\tera\hertz}$ and $\SI{190}{\tera\hertz}$.
Considering the Lorentz model~\eqref{eqlorentzfreq} we note that the phase
matching condition for optical paramectric oscillation is almost fulfilled for the
process of frequency splitting with the pump frequencies $\omega_{1}$ and
$\omega_{2}$, idler frequency $\omega_{i}=\SI{100}{\tera\hertz}$ and respective
signal frequencies
$\omega_{s,\,1}=\omega_{1}-\omega_{i},\:\omega_{s,\,2}=\omega_{2}-\omega_{i}$.

We conjecture that the significant amount of radiation at
$\SI{100}{\tera\hertz}$ and $\SI{190}{\tera\hertz}$ is due to the frequency
splitting described above. This hypotheses fits well with the simulation data,
where we observe a fast growth of
the conversion efficiency. Further, we conjecture
that the $\SI{400}{\tera\hertz}$ radiation is a result of frequency mixing processes of
the pump, signal and idler frequencies. This corresponds to the observation that
there is not a singular peak at one specific frequency but rather multiple
frequencies. In the last layers, we see a peak near $\SI{380}{\tera\hertz}$,
corresponding to the frequency doubling of the signal frequencies
$\omega_{s,\,1}$ and $\omega_{s,\,2}$. The radiation observed at
$\SI{400}{\tera\hertz}$ can be attributed to sum frequency generation of
$\omega_{i}$ and $\omega_{1}$ or $\omega_{2}$.

In Fig.~\ref{fig:freqplot2CE}~(b) we show the development of the internal
conversion efficiencies over the layers for the THz spectrum from
\SIrange{0.1}{0.5}{\tera\hertz}. We are able to observe a
growing THz conversion efficiency, since the process is phase-matched.
Furthermore, Fig.~\ref{fig:freqplot2CE}~(b) shows that the simulation data are in
good agreement with the experimental data obtained
from~\cite{olgunHighlyEfficientGeneration2022}, with a relative error of $9.8\,\%$
and an absolute error of $\num{0.0039}$.
Although more work is needed to
verify this beyond the results presented here, which are based on a single instance.
We interpolated the experimental
data between the start of the crystal with 0\% conversion efficiency and the end
of the crystal, where experimental data are available.

In Fig.~\ref{fig:freqplot2CE}~(c) we show the development of the internal
conversion efficiencies over the layers for the spectrum from
\SIrange{100}{800}{\tera\hertz}. Experimental data
for a comparison in this frequency range was not available. The decay of the
conversion efficiency over the
layers is physically plausible due to the the optical to \si{\tera\hertz}
conversion shown in Fig.~\ref{fig:freqplot2CE}~(b). We note
that the conversion efficiencies in Fig.~\ref{fig:freqplot2CE}~(b) and (c) don't necessarily add up to 100\%
since they don't cover the whole frequency spectrum and we include absorption in
the \si{\tera\hertz} range.

In Fig.~\ref{fig:freqplot2CE}~(d) we show the development of the internal
conversion efficiencies over the layers from
\SIrange{560}{600}{\tera\hertz}, i.\,e.\ the second harmonic. We see that
the conversion efficiency to the second harmonic doesn't grow significantly. A
linear regression reveals that the conversion efficiency increases slowly and
stays almost constant on average. The reason for the fluctuations is the
phase-mismatch, which leads to oscillating negative and positive interference. This
leads to varying conversion efficiencies over the layers, depending on how close
we are to a phase-match. This is in contrast to the \(\SI{0.3}{\tera\hertz}\)
radiation in Fig.~\ref{fig:freqplot2CE}~(b), where the processes are
phase-matched. It is all the more surprising that we generate so much of the
undesired radiation of the second harmonic frequency.

The results presented in Fig.~\ref{fig:freqplot2CE} will be used for the
optimization of the optical to \si{\tera\hertz} conversion, the validation of
simplified simulation algorithms and the development of multiscale models.
The
validation of the optical parametric oscillation responsible for the generation
of \SIlist{100;200;400}{\tera\hertz} radiation is subject to future work.
Frequency division in PPLN has been used in practice in the past;
cf.~\cite{gorelikCascadedOpticalParametric2006,neeOpticalFrequencyDivision1998}.
However, some of the effects observed in the simulations in this work still need
to be verified by experiments. In the current model we assumed the refractive
index in the optical region at frequencies larger $\SI{50}{\tera\hertz}$ as
constant. This is not correct and most likely leads to the parametric
amplification at 100 and 200 THz, which may not be present in reality. In order
to model this wavelength range correctly, the dispersion of the refractive index
in the optical wavelength range needs to be implemented. This is beyond the
scope of this paper, which is the development of a numerical algorithm for the
efficient solution of the nonlinear wave equation.

\FloatBarrier
\section{Conclusion}\label{sec:orgeacbc24}
In this work we have simulated THz generation in PPLN\@. To this
end we developed the physical model for nonlinear, dispersive, electromagnetic
wave propagation. We investigated the Lorentz dispersion model and absorbing
boundary conditions where we derived the governing
equations~\ref{def:lor-ade-cfs-pml}. We have demonstrated the efficiency of the
simulation methods and absorbing boundary conditions that
we implemented with the PML. They fit well into the STFEM framework, are effective and
increased the computational cost by only 2\%. We developed the methods in a
framework that is extendable to a wide range of applications within nonlinear
optics where numerical methods that are accurate and highly efficient are
needed.

In the future we want to extend our code to higher order methods within the
STFEM framework to further accelerate the simulations. The higher order methods
will allow us to chose larger time steps without the loss of accuracy. This is
crucial since simulation times can take up to 4 weeks right now. In order to not
overburden this presentation, we restricted ourselves to the 1D case. The methods and
implementation allow extensions to 2D and 3D and are subject to future work.

Our comparison with experimental data indicated good agreement in terms of the
main quantity of interest, the optical to \si{\tera\hertz} conversion
efficiency. The fact that we are able to simulate the THz conversion as well as
the harmonic generation will allow us to design better effective nonlinear media
by poling. We will use the results presented here to optimize the optical to
\si{\tera\hertz} conversion further. Future work include studies on this topic.
Furthermore a comparison with experimental data obtained from measurements in
the optical spectrum is needed to explain and verify the simulation results
further.
We may need to adjust the dispersion model in the optical
wavelength range. In the STFEM framework this can be done with little work and
low additional computational cost.

The simulations for the results we present here are computationally feasible but
still expensive. The main bottleneck is the time discretization. Since time
steps are inherently sequential, the time discretization limits the scalability
of parallelization.


\paragraph*{Acknowledgement}
NM acknowledges support by the Helmholtz-Gesellschaft grant number HIDSS-0002
DASHH. The authors would like to thank Prof.~Nina Rohringer for the helpful
discussions, especially regarding the interpretation of the simulation results.

\printbibliography




\appendix

\section{Derivation of the fully discrete system}\label{sec:app-fully-disc}
We derive the fully discrete system introduced in
Sec.~\ref{sec:linear-galerkin-method} in detail, especially the steps necessary
to get equation~\eqref{eqcg1} from~\eqref{eq:wavephy-variational}. The
representation of
\((e_{\tau,\,h},\,a_{\tau,\,h},\,p_{\tau,\,h},\,u_{\tau,\,h})\) in terms of
linear polynomials \(\phi_{n,0}\) and
\(\phi_{n,1}\in \mathbb{P}_{1}(\bar{I}_{n};\,\R)\) on \(\bar{I}_{n}\) can be
substituted into~\eqref{eq:wavephy-variational}. Here \(\phi_{n,0}\) and
\(\phi_{n,1}\) are the Lagrange polynomials on \(\bar{I}_{n}\) defined by the
conditions
\begin{equation}
  \label{}
  \phi_{n,0}(t_{n-1})=1,\quad \phi_{n,0}(t_{n})=0,\quad \phi_{n,1}(t_{n-1})=0,\quad \phi_{n,0}(t_{n})=1\,.
\end{equation}
We represent the discrete solutions \(u_{\tau,\,h}\), \(p_{\tau,\,h}\),
\(a_{\tau,\,h}\) and \(e_{\tau,\,h}\) on \(I_{n}\) in terms of the basis
functions as
\begin{equation}
  \begin{aligned}
    \label{eq:basis-rep}
    u_{\tau,\,h}(t)&=u_{h}^{n-1}\varphi_{n,\,0}(t)+u_{h}^{n}\varphi_{n,\,1}(t), &p_{\tau,\,h}(t)&=p_{h}^{n-1}\varphi_{n,\,0}(t)+p_{h}^{n}\varphi_{n,\,1}(t),\\
    a_{\tau,\,h}(t)&=a_{h}^{n-1}\varphi_{n,\,0}(t)+a_{h}^{n}\varphi_{n,\,1}(t), &e_{\tau,\,h}(t)&=e_{h}^{n-1}\varphi_{n,\,0}(t)+e_{h}^{n}\varphi_{n,\,1}(t),
  \end{aligned}
\end{equation}
with coefficients
\(u_{h}^{n-1},\: u_{h}^{n},\:p_{h}^{n-1},\: p_{h}^{n},\: a_{h}^{n-1},\:
a_{h}^{n}\in \mathcal{V}^{1}_{h}\) and
\(e_{h}^{n-1},\:e_{h}^{n}\in \mathcal{V}^{0}_{h}\). They are given by
\begin{equation}
  \begin{aligned}
    u^{n-1}_{h} &= u_{\tau,\,h}^{+}(t_{n-1}),& u^{n}_{h}&=u_{\tau,\,h}(t_{n}),& p^{n-1}_{h} &= p_{\tau,\,h}^{+}(t_{n-1}),& p^{n}_{h}&=p_{\tau,\,h}(t_{n}),\\
    a^{n-1}_{h} &= a_{\tau,\,h}^{+}(t_{n-1}),& a^{n}_{h}&=a_{\tau,\,h}(t_{n}),& e^{n-1}_{h} &= e_{\tau,\,h}^{+}(t_{n-1}),& e^{n}_{h}&=e_{\tau,\,h}(t_{n})\,.
  \end{aligned}
\end{equation}
Now we substitute~\eqref{eq:basis-rep} into~\eqref{eq:wavephy-variational} and
evaluate the time integrals numerically. We chose the two point Gauß-Lobatto
quadrature since it integrates linear polynomials exactly and is therefore exact
for the integrals at the left-hand side of~\eqref{eq:wavephy-variational}. This
lets us~\eqref{eq:wavephy-variational} into a time marching process over all
\(N\) subintervals with an initial value
\(e_{\tau,\,h}=e_{0,\,h}\in \mathcal{V}^{0}_{h}\). We note that the nonlinearity
does not require special treatment in this case, as can be seen by exact
integration of~\eqref{eq:wavephy-variational-c}. From the evaluation of the time
integrals we arrive at the fully disrete problem~\eqref{eqcg1}.

\section{Condensed systems of equation}\label{sec:app-condense-sys}
In the physical region we can condense the system equations such that we only
need to solve for \(e_{h}^{n}\). The other variables can be solved by cheap and
simple vector updates. This condenstation yields
\begin{subequations}
  \begin{multline}
    \label{cgp1-e}
    \paran{n_{\omega}^{2}+\frac{(n_{\Omega}^{2}-n_{\omega}^{2}) k^2
        \nu_{t}^2}{k^2 \nu_{t}^2+2 \Gamma_{0} k+4}}\mat{M}e_{h}^{n}
    + \frac{k^{2}}{4}\mat{A}e_{h}^{n} + \mat{N}(e_{h}^{n}) e_{h}^{n}\\
    =\frac{k^{2}}{4} \paran{\vct{b}_{h}^{n-1} + \vct{b}_{h}^{n}} +\frac{k^3
      \nu_{t}^2+2 \Gamma_{0} k^2}{k^2 \nu_{t}^2+2 \Gamma_{0}
      k+4}\mat{M}u_{h}^{n-1} +\frac{2 k^2 \nu_{t}^2+4 \Gamma_{0} k}{k^2
      \nu_{t}^2+2
      \Gamma_{0} k+4}\mat{M}p_{h}^{n-1}\\
    +\paran{n_{\omega}^{2}-\frac{(n_{\Omega}^{2}-n_{\omega}^{2}) k^2
        \nu_{t}^2}{k^2 \nu_{t}^2+2 \Gamma_{0} k+4}}\mat{M}e_{h}^{n-1} -
    \frac{k^{2}}{4}\mat{A}e_{h}^{n-1}+ \mat{N}(e_{h}^{n-1}) e_{h}^{n-1} +
    k\mat{M}a^0\,,
  \end{multline}
  \begin{align}
    \label{cgp1-p}
    \notag p_{h}^{n}&= \frac{4 k}{k^2 \nu_{t}^2+2 \Gamma_{0} k+4}u_{h}^{n-1} - \frac{k^2 \nu_{t}^2+2 \Gamma_{0} k-4}{k^2
                      \nu_{t}^2+2 \Gamma_{0} k+4}p_{h}^{n-1}\\
                    &\qquad+\frac{(n_{\Omega}^{2}-n_{\omega}^{2}) k^2
                      \nu_{t}^2}{k^2 \nu_{t}^2+2 \Gamma_{0} k+4}e_{h}^{n-1} +\frac{(n_{\Omega}^{2}-n_{\omega}^{2})
                      k^2 \nu_{t}^2}{k^2 \nu_{t}^2+2 \Gamma_{0} k+4}e_{h}^{n}\,,\\
    \label{cgp1-u}
    u_{h}^{n}&= -u_{h}^{n-1}+\frac{\Gamma_{0} k-2}{k} p_{h}^{n-1} + \frac{\Gamma_{0} k+2}{k} p_{h}^{n}\,,\\
    \label{cgp1-a}
    \notag \mat{M}a_{h}^{n}&= -\Gamma p_{h}^{n-1} - \frac{2n_{\omega}^{2}}{k}\mat{M}e_{h}^{n-1} -
                             \frac{2\chi^{(2)}}{k}\mat{N}(e_{h}^{n-1})e_{h}^{n-1} - \mat{M}a_{h}^{n-1}\\
                    &\qquad-\Gamma_{h}^{n-1} p_{h}^{n}
                      + \frac{2n_{\omega}^{2}}{k}\mat{M}e_{h}^{n} +
                      \frac{2\chi^{(2)}}{k}\mat{N}(e_{h}^{n})e_{h}^{n}\,.
  \end{align}
\end{subequations}
In a post processing step the equations~\eqref{cgp1-p},~\eqref{cgp1-u}
and~\eqref{cgp1-a} can then be solved as simple vector updates. We are only able
to compute \(a_{h}^{n}\) in the integrated form, since we have to include the
nonlinearity in the update for \(a_{h}^{n}\); cf.~\eqref{cgp1-a}. We could solve
for \(a_{h}^{n}\) by inverting a mass matrix. As can be seen in~\eqref{cgp1-e},
the term \(a_{h}^{n}\) is needed to assemble the right hand side, therefore this
is not necessary.

Inside the PML region the operators are of different form and we have the
auxiliary variables \(Q\) and \(R\). It is still possible to condense the system
of equations to the variable \(E\) and derive vector updates for the other
variables. To simplify the notation we introduce the two functions
\begin{subequations}
  \begin{align}
    \notag\mu_{e_{1}}&=\frac{
                       \alpha \kappa (n_{\omega}^{2} (k^3 vt^2+2 T k^2+4 k)+(n_{\Omega}^{2}-n_{\omega}^{2}) k^3 vt^2)}
                       {(\alpha k+2) (k^2 vt^2+2 T k+4)} + \frac{n_{\omega}^{2} \sigma k}
                       {(\alpha k+2)}\\
                     &\quad + \frac{
                       \kappa
                       (n_{\omega}^{2} (2 k^2 vt^2+4 T k+8)+2 (n_{\Omega}^{2}-n_{\omega}^{2}) k^2 vt^2)}
                       {(\alpha k+2) (k^2 vt^2+2 T k+4)}\,,\\
    \notag\mu_{e_{0}}&= \frac{
                       \alpha \kappa (n_{\omega}^{2} (k^3 vt^2+2 T k^2+4 k)- (n_{\Omega}^{2}-n_{\omega}^{2}) k^3
                       vt^2)}
                       {(\alpha k+2) (k^2 vt^2+2 T k+4)}-\frac{n_{\omega}^{2} \sigma k}{(\alpha k+2)}\\
                     &\quad + \frac{
                       \kappa (n_{\omega}^{2} (2 k^2 vt^2+4 T k+8)-2 (n_{\Omega}^{2}-n_{\omega}^{2}) k^2
                       vt^2)}
                       {(\alpha k+2) (k^2 vt^2+2 T k+4)}\,.
  \end{align}
\end{subequations}
With this we formulate the linear system of equations as
\begin{subequations}
  \begin{align}
    \label{cgp1-pml}
    \notag
    \mu_{e_{1}}
    \mat{M}e_{h}^{n} &- \frac{k^2 (\tilde{\sigma} k \kappa-\tilde{\alpha} k-2)}
                       {4 (\tilde{\alpha} k+2) \kappa}\mat{A}\\
    \notag
    \quad&=
           -\frac{ k^2}{\tilde{\alpha} k+2}\mat{G}q_{h}^{n-1}
           \frac{2 \alpha k}{\alpha k+2}\mat{M}r_{h}^{n-1}
           \frac{k^2 (k vt^2+2 T)}{k^2 vt^2+2 T k+4}\mat{M}u_{h}^{n-1}
           \frac{2 k (k vt^2+2 T)}{k^2 vt^2+2 T k+4}\mat{M}p_{h}^{n-1}\\
    \quad &+\mu_{e_{0}}\mat{M}e_{h}^{n-1}+ \frac{k^2 (\tilde{\sigma} k \kappa- \tilde{\alpha} k-2)}
            {4(\tilde{\alpha} k+2) \kappa}\mat{A}e_{h}^{n-1}
            +k\mat{M}a_{h}^{n-1}\,,\\
    \label{cgp1-pml-p}
    \notag
    p_{h}^{n}&=\frac{4k}{k^2\nu_{t}^2+2\Gamma_0k+4}u_{h}^{n-1}
               -\frac{k^2\nu_{t}^2+2\Gamma_0k-4}{k^2\nu_{t}^2+2\Gamma_0k+4}p_{h}^{n-1}\\
                     &\quad+\frac{(n_{\Omega}^{2}-n_{\omega}^{2})k^2\kappa\nu_{t}^2}{k^2\nu_{t}^2+2\Gamma_0k+4}e_{h}^{n-1}
                       +\frac{(n_{\Omega}^{2}-n_{\omega}^{2})k^2\kappa\nu_{t}^2}{k^2\nu_{t}^2+2\Gamma_0k+4}e_{h}^{n}\,,\\
    \label{cgp1-pml-q}
    \mat{M}q_{h}^{n}&=-\frac{\tilde{\alpha}k-2}{\tilde{\alpha}k+2}q_{h}^{n-1}
                      -\frac{\tilde{\sigma}k}{(\tilde{\alpha}k+2)}\mat{A}e_{h}^{n-1}
                      -\frac{\tilde{\sigma}k}{(\tilde{\alpha}k+2)}\mat{A}e_{h}^{n}\,,\\
    \label{cgp1-pml-u}
    u_{h}^{n}&= -u_{h}^{n-1}+\frac{\Gamma_{0} k-2}{k} p_{h}^{n-1}
               + \frac{\Gamma_{0} k+2}{k}p_{h}^{n}\,,\\
    \label{cgp1-pml-r}
    r_{h}^{n}&=-\frac{\alpha k-2}{\alpha k+2}r_{h}^{n-1}
               + \frac{n_{\omega}^{2}\sigma k}{\alpha k+2}e_{h}^{n-1}
               + \frac{n_{\omega}^{2}\sigma k}{\alpha k+2}e_{h}^{n}\,,\\
    \label{cgp1-pml-a}
    \notag
    a_{h}^{n}&=-\alpha r_{h}^{n-1}
               -\Gamma_{0}p_{h}^{n-1}
               + \frac{n_{\omega}^{2}(\sigma k-2\kappa)}{k} e_{h}^{n-1}
               -a_{h}^{n-1}\\
                     &\quad-\alpha r_{h}^{n}
                       -\Gamma_{0}p_{h}^{n}
                       + \frac{n_{\omega}^{2}(\sigma k+2\kappa)}{k} e_{h}^{n}\,.
  \end{align}
\end{subequations}
The update equation~\eqref{cgp1-pml-q} can only be solved by inverting a mass
matrix, since we need the gradient of \(q_{h}^{n}\) in~\eqref{cgp1-pml} for the
assembly of the right-hand side. However, the computations necessary for the
solution of the variables \(q_{h}^{n}\) and \(r_{h}^{n}\) can be restricted to
the PML domain. Therefore, depending on the size of the PML compared to the
physical domain, even the mass-matrix inversion has shown to be neglibile within
the whole solution process.
\end{document}